\documentstyle[preprint,aps,epsfig]{revtex}
\begin{document}
\draft
\preprint{ SNUTP 96-061} 
\bigskip
\title{ Color-octet heavy quarkonium productions \\ in  
$Z^{0}$ decays at LEP }
\author{Seungwon Baek\thanks{ swbaek@phya.snu.ac.kr}$^{(a)}$, 
P. Ko\thanks{ pko@phyb.snu.ac.kr}$^{(b)}$ ,
Jungil Lee\thanks{ jungil@phya.snu.ac.kr}$^{(a)}$, and 
H.S. Song\thanks{ hssong@physs.snu.ac.kr}$^{(a)} $}
\address{$^{(a)}$ 
Center for Theoretical Physics and
Department of Physics,\\ Seoul National University, 
Seoul 151-742, Korea \\
$^{(b)}$ 
Department of Physics, Hong-Ik University, Seoul 121-791, Korea
}
\date{\today}
\maketitle
\begin{abstract}
We consider the energy and the polar angle distributions of $J/\psi$'s 
produced via the color-singlet and the color-octet (${^3S_1^{(8)}}$) 
mechanisms in $Z^0 \rightarrow  J/\psi +X$ at LEP.
Since both distributions  of the $J/\psi$ 
produced via color-octet mechanism are  
significantly different from those via color-singlet mechanism,  these
observables can be used as tests of color-octet production 
mechanism for heavy   quarkonia. 
We also discuss $Z^{0} \rightarrow \Upsilon + X$ and $W \rightarrow J/\psi
({\rm or \Upsilon}) + X$ in brief. 
\end{abstract}


\narrowtext


Heavy quarkonium productions in high energy processes have long been 
studied in perturbative QCD along with the color-singlet model.
However, the recent measurements of $J/\psi$ and $\psi^{'}$ productions at 
the Tevatron put serious questions on such theoretical approaches 
\cite{mangano}. 
In short, the data for $p \bar{p} \rightarrow \psi^{'} + X$ is larger
than the color-singlet model by a factor of $\sim 30$. 
In order to resolve this discrepancy, Braaten and Fleming suggested a 
color-octet fragmentation of an energetic gluon into $\psi^{'}$ by 
emission of soft gluons \cite{fleming}.  
This approach requires one parameter, $\langle 
0 | O_{8}^{\psi^{'}} (^{3}S_{1}) | 0 \rangle$, the 
matrix element of a color-octet, dimension-six, four-quark operator in 
the Nonrelativistic QCD, which is the effective field theory for QCD
relevant to the heavy quarkonium physics \cite{lepage}.  
Braaten and Fleming showed that if one fixes this parameter to fit
the measured cross section, then the observed spectrum in $p_T$ is 
reproduced.  Then, Cho and Leibovich \cite{cho1} extended this approach 
in order to go beyond the fragmentation picture 
(which is applicable only  to the limit $p_{T}^{2} \gg m_{\psi^{'}}^2$),
including other color-octet intermediate states such as ${^{1}S_{0}}^{(8)}$ 
and ${^{3}P_{J=0,1,2}}^{(8)}$ in terms of two more NRQCD matrix elements,
$\langle 0 | O_{8}^{\psi^{'}} (^{1}S_{0}) | 0 \rangle$ and 
$\langle 0 | O_{8}^{\psi^{'}} (^{3}P_{0}) | 0 \rangle$.
This is especially relevant 
for the case of inclusive $\Upsilon (nS)$ productions  at the Tevatron, 
since the $p_T$ in this case is not that large compared to $m_{\Upsilon}$
and thus the fragmentation picture is not valid any more.  
Overall descriptions for $\psi^{(')}$ and $\Upsilon (nS)$ productions
at the Tevatron seem reasonably good by now, although some color-octet matrix 
elements are shown to lead  too large contributions to $J/\psi$ 
photoproductions \cite{photo} and $B \rightarrow J/\psi + X$ \cite{ko2}.  

In view of this, it is important to test the idea of color-octet mechanism
in the heavy quarkonium productions in places other than at the Tevatron.
One can think of the $S-$wave charmonium production in $B$ decays 
\cite{ko2,ko1}, 
the angular dependence 
of $J/\psi$ production cross section in the $e^{+} e^{-}$ 
annihilations \cite{chen}, $Z^{0} \rightarrow J/\psi + X$ \cite{keung,cho2},
$J/\psi$ productions at the $\gamma p$ collision \cite{photo,ko2},
the associated $J/\psi + \gamma$ hadroproduction \cite{jungil}
and many others. All of these have formed an active research field recently 
\cite{annual}.

In this letter, we reconsider the inclusive heavy quarkonium production 
in $Z^{0}$ decays, which is accessible at the current LEP experiments.
This study is partly motivated by the recent report by OPAL collaboration
\cite{opal}  
that they have observed excess of events for $Z^{0} \rightarrow \Upsilon (nS)
+ X$ for $n=1,2,3$, 
larger than theoretical expectation by a factor of $\sim 10$,
\begin{equation}
\sum_{n=1}^{3} B(Z^{0} \rightarrow \Upsilon (nS) + X)_{\rm exp} 
\simeq (1.2_{-0.6}^{+0.9} \pm 0.2) \times 10^{-4}, 
\end{equation}
compared to the $b-$quark fragmentation contribution \cite{yuan},
\begin{equation}
\sum_{n=1}^{3} B(Z^{0} \rightarrow \Upsilon (nS) + X)_{\rm frag} 
\simeq  1.6  \times 10^{-5}.
\end{equation}
Similar excess were also observed in the prompt $J/\psi$ and $\psi^{'}$
production in $Z^0$ decays, although the experimental errors are quite
large \cite{opal2} :
\begin{eqnarray}
B(Z^{0} \rightarrow J/\psi + X) & = & (3.0 \pm 2.3) \times 10^{-4},
~~~~~( 2.6 \times 10^{-5})
\\
B(Z^{0} \rightarrow \psi^{'} + X) & = & (2.2 \pm 1.5) \times 10^{-4},
~~~~~( 1.1 \times 10^{-5})
\end{eqnarray}
where the results for the heavy quark fragmentations \cite{yuan} 
are shown in the parentheses for comparison.
The issues of the branching ratios and the energy distributions of 
$J/\psi$ in $Z^0 \rightarrow J/\psi +X$ have been addressed already in 
Refs.~\cite{keung} and \cite{cho2}.  
In this letter, we suggest another observable,
the polar angle distribution of $J/\psi$ (relative to the $e^{+} e^{-}$ beam  
directions at LEP) for the color-singlet and 
the color-octet contributions,  in order to provide another independent 
check of  the idea of color-octet mechanism in heavy quarkonium productions. 

In order to study the angular distribution of $J/\psi$ produced in 
the $Z^{0}$ decays at LEP, it  is convenient to define 
$S(E)$ and $\alpha (E)$ as 
\begin{equation}
{d^{2} \Gamma_{1,8} \over dE_{\psi} d\cos \theta_{\psi}} \equiv
S_{1,8} (E_{\psi})~(1 + \alpha_{1,8} (E_{\psi})~ \cos^2 \theta_{\psi} ),
\end{equation}
following the case of $e^{+} e^{-} \rightarrow \gamma^{*} \rightarrow
J/\psi + X$ at CLEO energy, $\sqrt{s} \approx 10.6 $ GeV \cite{chen,cho3}.
The subscripts $1,8$ denote the color-singlet and the color-octet 
contributions, respectively.  
Since we are concerned with the polar angle ($\cos \theta_{\psi}$)
distribution of $J/\psi$ in the
rest frame of $Z^0$ as well as the energy distribution, we have to include
the polarization of the $Z^0$ produced through the $e^+ e^-$
annihilations at LEP (with $\sqrt{s} = M_Z$).
This effect  can be conveniently described
by replacement of $\epsilon^{\mu} (Z) \epsilon^{*\nu} (Z)$ in the decay
process by the density matrix $\rho_Z^{\mu\nu}$,

\begin{equation}
\rho_Z^{\mu\nu}  =  {1 \over 3}
\left(- g^{\mu\nu} + {Z^{\mu} Z^{\nu}\over M_Z^2}\right)
       -\frac{i}{2M_Z}\epsilon^{\mu\nu\lambda\tau}Z_\lambda{\cal P}_\tau
                 -{1 \over 2} {\cal Q}^{\mu\nu},
\label{eq:pol}
\end{equation}
where $Z^\mu$ is the 4-momentum of $Z^0$.

The vector polarization ${\cal P}^\mu$ and the tensor polarization 
${\cal Q}^{\mu\nu }$
in Eq.~(\ref{eq:pol}) can be obtained in the case of 
$e^{+} (k_2) ~e^{-}(k_1) \rightarrow Z^{0}$
process as following \cite{song}~:

\begin{eqnarray}
{\cal P}^{\mu} & = & {\Delta^{\mu} \over M_Z}
   ~~\frac{2  g_V^e g_A^e}{(g_V^e)^2 + (g_A^e)^2},
\nonumber
\\
{\cal Q}^{\mu \nu} & = & 
-{1 \over 3} \left(- g^{\mu\nu} + {Z^{\mu} Z^{\nu}\over M_Z^2}\right) 
+\frac{\Delta^\mu\Delta^\nu}{M_Z^2},
\end{eqnarray}
where $\Delta^\mu \equiv (k_1 -k_2)^\mu$.

After obtaining the invariant amplitudes for the color-singlet and the 
color-octet processes, we have performed the symbolic manipulations over  
the squared amplitude with the above polarization tensors 
and integrated over the phase space, using REDUCE and Mathematica.
The resulting expressions for $S_{1,8}(E)$ and 
$\alpha_{1,8}(E)$ are rather lengthy, and will be shown separately elsewhere
\cite{song1}.  Our formalism with the density matrices for polarized particles 
can be applied to other 
processes where $Z^0$'s are produced via different mechanism from the 
$e^{+} e^{-} \rightarrow Z^0$.  For such processes, the density matrix for
$Z^0$ will be different from Eqs.~(7).  The explicit relation between 
the density matrices of
$J/\psi$ and $Z^0$ in $Z^{0} \rightarrow J/\psi +X$ for an arbitrary $Z^0$ 
polarization  will be discussed there, too.

Let us first consider the conventional color-singlet contribution to $Z^0 
\rightarrow J/\psi +X$ at LEP. The most dominant contribution at LEP 
energy scale is known \cite{keung,cho2} to come from $Z^{0} 
\rightarrow (c \bar{c}) ({^{3}S_{1}^{(1)}})
+ c + \bar{c}$, which can be further approximated into the (anti)$c-$quark
fragmentation into $J/\psi$ as \cite{yuan}.  In this work, we employ the      
full amplitude for the color-singlet $J/\psi$ production without using the
fragmentation approximation.  The relevant  Feynman diagrams are shown in  
Fig.~1. The corresponding amplitude can be obtained as usual in the covariant
gauge.  For simplicity, we do not show the amplitude 
for the color-singlet $J/\psi$ production in $Z^{0} \rightarrow J/\psi +X$ 
explicitly.  However, the corresponding expressions for $S_1 (E)$ and 
$\alpha_1 (E)$ will be given later \cite{song1}.

Next, we consider $Z^{0} \rightarrow q \bar{q} + g$ followed by
$g \rightarrow {^{3}S_{1}}^{(8)} \rightarrow J/\psi +$(soft hadrons) 
with $q = u,d,c,s,b$. The relevant Feynman diagrams are shown in Fig.~2.
Other diagrams are higher orders in $v^2$, or suppressed by the short
distance factor $m_{Q}^{2} / M_{Z}^2$,  and thus irrelevant to
$Z^{0} \rightarrow q \bar{q} + J/\psi (~{\rm or} \psi^{'}, \Upsilon (nS))$.
It is convenient to use the following variables :
\begin{eqnarray}
s & \equiv & ( Z - P )^{2} = ( q_1+q_2 )^2, \\
t & \equiv & ( Z - q_1 )^{2} = ( P + q_2)^2, \\
u & \equiv & ( Z - q_2 )^{2} = ( P + q_1 )^2,
\end{eqnarray}
with $s + t + u = M_Z^{2} + M_{\psi}^2+2m_q^2$,
and $P$, $q_1$ and $q_2$ are the 4-momenta of $J/\psi$,
$q$ and $\overline q$, respectively.  
The color-octet contribution of $J/\psi$ production
via the process $Z_0\rightarrow q\bar{q} + J/\psi$
factorizes into short and long distance pieces as
\begin{eqnarray}
&{\cal M}&(Z_0\rightarrow q\bar{q} J/\psi )_{\mbox{\tiny octet}}
\nonumber\\
&=&{\cal M}(Z_0\rightarrow q\bar{q} g^*(\rightarrow c\bar{c}[^3S_1^{(8)}]))
_{\mbox{\tiny short distance}}
\times
{\cal M}(c\bar{c}[^3S_1^{(8)}]\rightarrow J/\psi)
_{\mbox{\tiny long distance}}.
\end{eqnarray}
In terms of $s,t,u$ variables, the amplitude for the process 
$Z_0\rightarrow q\bar{q} g^*(\rightarrow c\bar{c}[^3S_1^{(8)}])$ 
is given by  
\begin{equation}
{\cal M}(Z_0\rightarrow q\bar{q} g^*(\rightarrow c\bar{c}[^3S_1^{(8)}])_{
\rm octet}  
=\frac{eg_s^2}{M_{J/\psi} \sin\theta_W\cos\theta_W}
({\cal M}_1^{\mu\nu}+{\cal M}_2^{\mu\nu})
\epsilon_\mu^Z(Z)
\epsilon_\nu^{*\psi}(P),
\end{equation}
\begin{eqnarray}
{\cal M}_1^{\mu\nu}&=&
\bar{u}(q_1)T_a\gamma^\nu\frac{\not{P}+\not{q}_1+m_q}{u-m_q^2}
\gamma^\mu(g_V+g_A\gamma_5)v(q_2),\nonumber\\
{\cal M}_2^{\mu\nu}&=&
\bar{u}(q_1)T_a\gamma^\mu(g_V+g_A\gamma_5)
\frac{\not{q}_1-\not{Z}+m_q}{t-m_q^2}\gamma^\nu v(q_2).
\end{eqnarray}
where $g_V$ and $g_A$ for  the up-type and the down-type quarks are given as 
\begin{eqnarray}
(g_V,g_A)=\left\{
\begin{array}{cc}
\left(+\frac{1}{4}-\frac{2}{3}\sin^2\theta_W, -\frac{1}{4}\right),
&~~~{\rm for} 
~q=u,~c,~t,\\
\left(-\frac{1}{4}+\frac{1}{3}\sin^2\theta_W, +\frac{1}{4}\right),
&~~~{\rm for} ~q=d,~s,~b.
\end{array}
\right.
\end{eqnarray}
The long distance contribution due to the color-octet fragmentation of
a gluon into the color-singlet heavy quarkonium $J/\psi$, $\psi^{'}$ or 
$\Upsilon (nS)$  is described in terms of a parameter ${\cal M}_{8} (nS)$, 
which depends on the heavy quarkonium state one considers.
Cho and Leibovich have determined these parameters for $nS = \Upsilon (nS),
J/\psi$   and $\psi^{'}$ \cite{cho1,cho2} : for example, 
\begin{eqnarray}
\sum_{n=1}^{3}{\cal M}_8(\Upsilon(nS))&\equiv&  
\sum_{n=1}^{3}
|{\cal M}(b\bar{b}[^3S_1^{(8)}]\rightarrow \Upsilon(nS))|^2
 =  6.4\times 10^{-3}~{\rm GeV}^2, \\
{\cal M}_8(J/\psi(1S))&\equiv&
|{\cal M}(c\bar{c}[^3S_1^{(8)}]\rightarrow J/\psi)|^2
 =  0.68\times 10^{-3}~{\rm GeV}^2.
\end{eqnarray}
After one averages and sums the squared amplitude over the initial 
and the  final spin states using Eqs.~(6),  and performs the
phase space integration, one easily finds various distributions as well as 
the branching ratios.  

We first show the $z (\equiv 2 E_{\psi} / M_{Z})$ dependence of $\alpha_{1,8} 
(E_{\psi})$ for 
the color-singlet (in dashed curve) and the color-octet (in solid curve) 
case respectively in Fig.~3 (a).  We note that the $\alpha_{1}(E)$ abruptly  
changes its sign from $\sim - 0.8$ to $+1$ near the end point of $z \approx 
1$. This originates from the smallness of $\delta = 2M_{\psi} / M_Z \approx
6.6 \times 10^{-2}$.
\footnote{ For the case of the $J/\psi$ production at CLEO energy through 
$e^+ e^- \rightarrow 
\gamma^* \rightarrow J/\psi +X$ \cite{chen} \cite{cho3}, 
the corresponding $\delta = M_{\psi} / 
E = 3.1 ~{\rm GeV} / 5.29~{\rm GeV} \approx 0.59$ is not small.  }
In Fig.~3 (b). we show the $\alpha (E_{\Upsilon})$ for the case of 
$Z^{0} \rightarrow \Upsilon (1S) + X$. The behavior near $z \approx 1$ 
becomes less abrupt here, as a result of the not-too-small 
$\delta \approx 0.42$ in this case. 
We have checked that this abrupt behavior near $z \approx 1$ dies out 
as we increase $\delta$. 

Our energy distributions shown in Fig.~4 (a) agree with 
those obtained by  \cite{keung} and \cite{cho2}, which is a check of our 
calculations based on the polarized $Z^0$ decays.  
The color-octet mechanism (shown in the solid curve) produces
softer $J/\psi$'s compared to the color-singlet mechanism (shown in 
the dashed curve).  
Most $J/\psi$'s are predicted to have energy around $\sim 5.5$ GeV, if the 
color-octet mechanism works. Similar is true of the $\Upsilon$  energy 
distribution shown in Fig.~ 4 (b).  

The $\cos \theta_{\psi}$  distributions($\theta_\psi$ is the angle
between the initial electron and the final $J/\psi$ direction) can be 
obtained  by integrating (5) over the heavy quarkonium energy $E_{\psi}$.
The color-singlet contribution 
predicts the polar angle distribution to be
\begin{eqnarray}
{d \Gamma_{1} \over d\cos \theta_{\psi}} &=& 
  (5.4 \times 10^{-2})~(1 + 0.92 \cos^2\theta) ~{\rm MeV}  .
\\ 
{d \Gamma_{1} \over d\cos \theta_{\Upsilon}} &=& 
  (9.7 \times 10^{-3})~(1 + 0.73 \cos^2\theta) ~{\rm MeV}  .
\end{eqnarray}
On the other hand, the color-octet mechanism predicts
\begin{eqnarray}
{d \Gamma_{8} \over d\cos \theta_{\psi}} &=&    
  (25.8 \times 10^{-2})~(1 + 0.34 \cos^2\theta) ~{\rm MeV}  .
\\
{d \Gamma_{8} \over d\cos \theta_{\Upsilon}} &=&   
  (36.7 \times 10^{-3})~(1 + 0.27 \cos^2\theta) ~{\rm MeV}  .
\end{eqnarray}
Adding these two contributions together, we get 
\begin{eqnarray}
{d  \Gamma_{1+8}  \over d\cos \theta_{\psi}} &=&    
  (31.2 \times 10^{-2})~(1 + 0.44 \cos^2\theta) ~{\rm MeV}  .
\\ 
{d  \Gamma_{1+8}  \over d\cos \theta_{\Upsilon}} &=&   
  (46.4 \times 10^{-3})~(1 + 0.37 \cos^2\theta) ~{\rm MeV}  .
\end{eqnarray}
Therefore, we find that the  color-octet  mechanism makes the  angular 
distributions flatter compared to that by the color-singlet mechanism alone.
The polar angle distributions for $Z \rightarrow J/\psi + X$ and 
$Z \rightarrow \Upsilon + X$ are shown in Figs.~5 (a) and (b).
For each plot, the color-singlet, the color-octet and the sum of the two 
are shown in the dotted, the dashed and the solid curves, respectively.
The difference in the $\theta_{\psi}$ distributions between the singlet 
and the octet contributions are not so prominent compared with
the energy ($E_{\psi}$) distributions shown in Figs.~4 (a) and (b).
However, using the number of $\sim 600 ~J/\psi$'s from 
$Z^0 \rightarrow J/\psi + X$
at LEP, it would be possible to distinguish (21) from (17) from the 
$\cos \theta_{\psi}$ distribution \cite{dowon} (at least for the $J/\psi$ 
production in $Z^0$ decays).   Thus, the measurement of $\cos
\theta_{\psi}$  distribution could  constitute another independent test of 
the idea of color-octet mechanism as a possible solution to $\psi^{'}$ 
anomaly at the Tevatron.  Since the measurement of the polarization of 
$\psi^{'}$ is not easy, our option may be more viable experimentally.  

Finally, we can obtain the branching ratios for
$Z^{0} \rightarrow  J/\psi ({\rm or}~ \Upsilon (1S)) + X $
through the color-singlet and the color-octet mechanisms by integrating 
(17) --(23) over the whole $\theta_{\psi}$.
The results are shown in Table~1.
The color-octet contribution 
is comparable to that through the color-singlet contributions ( or 
equivalently, the  $c ({\rm or} ~b)-$quark fragmentation into 
$J/\psi ({\rm or} ~\Upsilon(1S))$ considered by Braaten, Cheung and Yuan
\cite{yuan}).  When summed over 
the light flavor quantum numbers ($q = u,d,c,s,b$), the former actually
becomes larger than the latter by a factor of $4-10$. 
For the $\psi^{'}$, $\Upsilon (2S)$ and $\Upsilon (3S)$
production rates, the color-octet
contributions should be scaled down by a factor of
$0.3$, $0.6$ and $0.14$, respectively, coming from
${\cal M}_{8}(2S) / {\cal M}_{8}(1S)$. Our results on the branching ratio 
agree with the results obtained in Refs.~\cite{keung} and \cite{cho2}, 
which is another check of our calculations based on polarized $Z^0$ decays.

Our analysis can be  easily extended to the $W$ decays,
$W \rightarrow q \bar{q}^{'} + g$ followed by $g \rightarrow J/\psi, 
({\rm or}~\psi^{'}, \Upsilon (nS))$.  Using the unitarity of the 
CKM matrix element $V_{qq^{'}}$, one can easily show that 
\begin{equation}
B( W \rightarrow J/\psi + X)_{\rm octet} \approx 2 \times 10^{-4},
\end{equation}
compared to the charm quark fragmentation in the singlet model \cite{yuan},
\begin{equation}
B(W \rightarrow J/\psi + X)_{\rm frag} \approx 4 \times 10^{-5}. 
\end{equation}  
Again, the color-octet mechanism dominates the heavy quark fragmentation
in the color-singlet model in $W$ decays, and this may be observed at
LEP200.  For Upsilon production, we have 
\begin{equation}
B( W \rightarrow \Upsilon (1S) + X)_{\rm octet}
 \approx 3.7 \times 10^{-5}.
\end{equation}

In conclusion, we have considered the color-octet mechanism in the
heavy quarkonium production in $Z^0$ decays. Both $Z^{0} \rightarrow 
\Upsilon + X$ and $J/\psi +X$ are enhanced due to the color-octet mechanism 
by a factor of $\sim 10$ or so.  Also, the energy and the polar angle 
distributions of $J/\psi$'s produced via color-octet mechanism are 
drastically different from those (via charm-quark fragmentation) in the
color-singlet model.  These two distributions can provide  key 
tests for the idea of the 
color-octet gluon fragmentation which was invented to solve the 
$\psi^{'}$ anomaly at the Tevatron. 
This idea might be the simultaneous solution to the
$\psi^{'}$ anomaly at the Tevatron and the excess of events in the channel
$Z^{0} \rightarrow \Upsilon +X$ at LEP, if the observed energy and polar 
angle distributions follow predictions made here and 
by other works \cite{keung}
\cite{cho2}.  
Similar phenomena occur in the case of 
inclusive $W$ decays into $J/\psi, \psi^{'}, \Upsilon (nS)$, and the excess
of $W \rightarrow \Upsilon + X$ would constitute another test of the
color-octet mechanism.

A remark is in order before closing.    Some recent 
works on the color-octet contributions to $J/\psi$ photoproduction 
\cite{photo,ko2}
and $B \rightarrow J/\psi +X$ \cite{ko1,ko2} 
show that some of the color-octet matrix
elements (via the ${^{1}S_{0}}^{(8)}$ and ${^{3}P_{J}}^{(8)}$ ) determined 
from the $J/\psi$ productions at the Tevatron \cite{cho1} might have been 
overestimated by an order of magnitude. However, in the $J/\psi$ productions
in $Z^0$ decays considered in this work, the color-octet ${^{3}S_{1}}^{(8)}$ 
channel is most dominant, with other color-octet channels being essentially 
negligible.  Therefore, our predictions would be relatively stable.

\acknowledgements
We thank Prof. Dowon Kim for discussions on the experimental
situations. 
This work was supported in part by KOSEF through CTP at Seoul National 
University, in part by Korea Research  Foundations,
and in part by the Basic Science Research
Program, Ministry of Education,  Project No. BSRI--96--2418.
P.K. is also supported in part by NON DIRECTED RESEARCH FUND, 
Korea Research  Foundations.


\newpage
\begin{center}
{FIGURE CAPTIONS}
\end{center}
\noindent
\vskip 1cm
Fig.1
\hskip .3cm
{
Feynman diagrams for  the color-singlet mechanism
for $Z^{0} \rightarrow (c \bar{c})({^{3}S_{1}^{(1)}}) + c \bar{c} $.
}
\\
\vskip 1cm
Fig.2
\hskip .3cm
{
Feynman diagrams for  the color-octet mechanism
for $Z^{0} \rightarrow q \bar{q} + J/\psi $ with  $q = u,d,c,s,b$.
}\\
\vskip 1cm
Fig.3
\hskip .3cm
{
 The $\alpha_{1,8} (z)$ of the (a) $J/\psi$ and (b) $\Upsilon$
 as functions of $z \equiv 2 E_H / M_Z~(H=\psi,\Upsilon)$
for the color-singlet (in dashed curve) and the color-octet (in solid
curve) case respectively.}
\\
\vskip 1cm
Fig.4
\hskip .3cm
{
The energy
distributions of the (a) $J/\psi$ and (b) $\Upsilon$.
The color-octet contribution  and the color-singlet contributions are shown
in the solid and the dashed curves, respectively.
}
\\
\vskip 1cm
Fig.5
\hskip .3cm
{
The polar angle
distributions of the (a) $J/\psi$ and (b) $\Upsilon$ (relative to
the $e^{+} e^{-}$ beam direction at the $Z-$peak).
The color-octet contribution, the color-singlet contributions and the sum 
of the two are shown
in the dashed, the dotted and the solid curves, respectively.
}
\begin{table}
\caption{ Branching ratios for $Z^{0} \rightarrow q \bar{q} + 
J/\psi ({\rm or} ~\Upsilon (1S) ) $ for $q=u,d,c,s,b$ with $m_{u} =
5 $ MeV, $m_{d} = 10 $ MeV, $m_{s} = 150 $ MeV, $m_{c} = 1.5 $ GeV and 
$m_{b} = 4.9 $ GeV ,and for the long-range matrix elements
$\sum_{n=1}^3{\cal M}_1 (\Upsilon(nS)) =2.3  \times 10^{-1} {\rm GeV^2}$,
$\sum_{n=1}^3{\cal M}_8 (\Upsilon(nS)) =6.4 \times 10^{-3} {\rm GeV^2}$,
${\cal M}_1 (J/\psi(1S)) =6.1 \times 10^{-2} {\rm GeV^2}$ and
${\cal M}_8 (J/\psi(1S)) =6.8 \times 10^{-4} {\rm GeV^2}$.
}
\label{table1}
\begin{tabular}{c|p{4.5cm}|p{4.5cm}}
$q$ & \hskip -.7cm $ B(Z^{0} \rightarrow q \bar{q} + J/\psi)$ & 
\hskip -.7cm $ B( Z^{0} \rightarrow
 q \bar{q} + \Upsilon (1S) ) $  
\\
$ u$  &  $  4.1 \times 10^{-5}$ & $ 5.6 \times 10^{-6}$
\\
$ d,s$  &  $  5.4 \times 10^{-5}$ & $ 7.2 \times 10^{-6}$
\\
$c $  & $ 3.9 \times 10^{-5}$ & $ 5.6 \times 10^{-6} $
\\
$b$      &  $ 4.4 \times 10^{-5}$ & $ 6.5 \times 10^{-6} $
\\
octet sum  &  $  23.2 \times 10^{-5}$ & $ 32.1 \times 10^{-6}$
\\
singlet & $5.6 \times 10^{-5}$ & $ 9.6 \times 10^{-6}$
\\
(fragmentation) & ($6.5 \times 10^{-5}$) & ($ 14.2 \times 10^{-6}$)
\\
total  & $28.8 \times 10^{-5}$  & $ 41.7 \times 10^{-6}$
\end{tabular}
\end{table}
\begin{figure}
\vskip .5cm
\hbox to\textwidth{\hss\epsfig{file=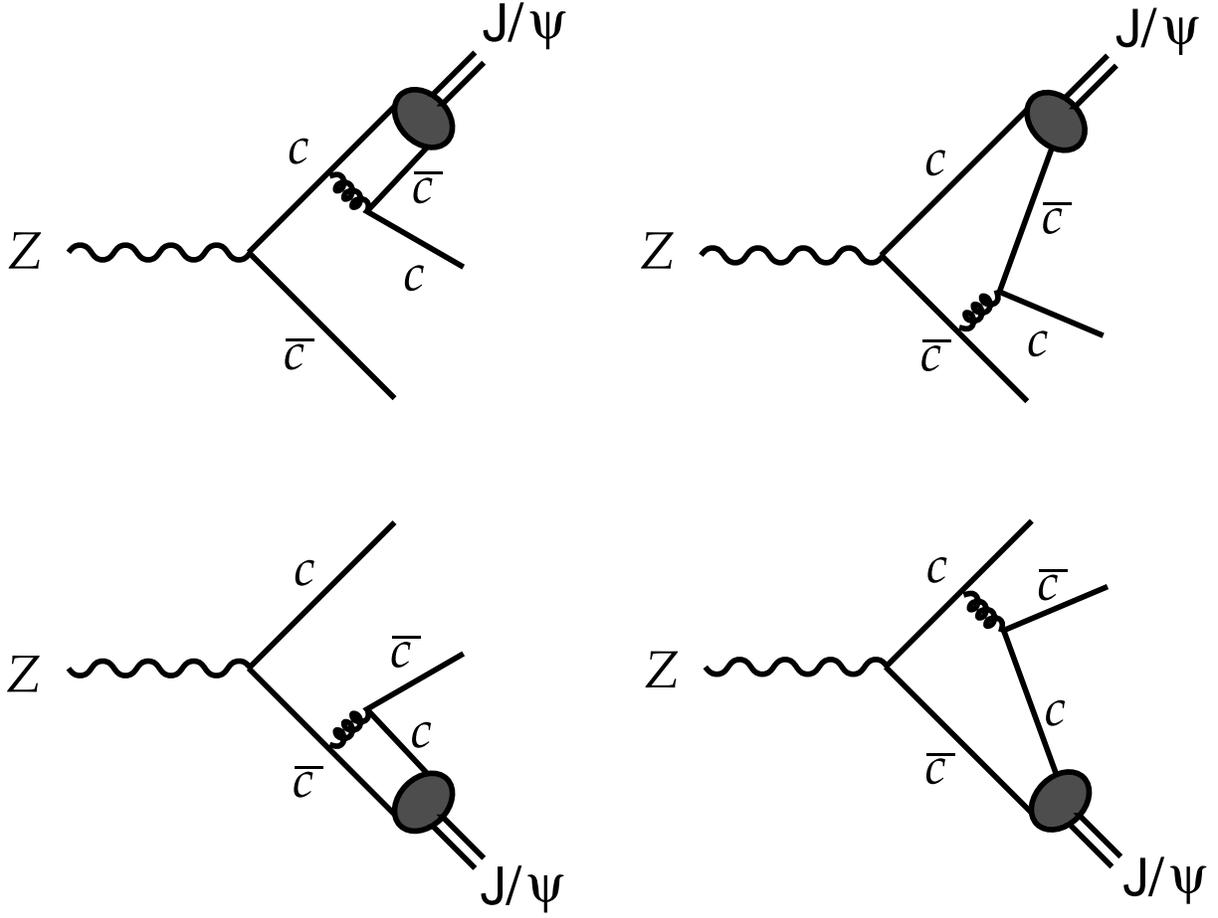,width=16cm}\hss}
\vskip 0.5cm
\caption{Feynman diagrams for  the color-singlet mechanism
for $Z^{0} \rightarrow (c \bar{c})({^{3}S_{1}^{(1)}}) + c \bar{c} $. 
}
\label{figone}
\end{figure}

\begin{figure}
\vskip .5cm
\hbox to\textwidth{\hss\epsfig{file=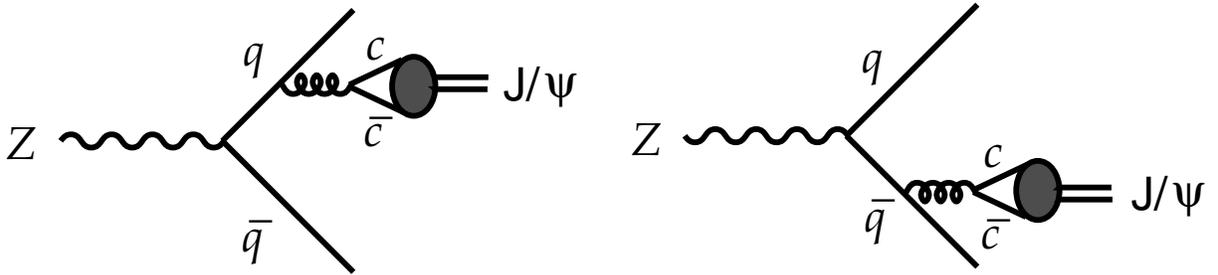,width=16cm}\hss}
\vskip 0.5cm
\caption{Feynman diagrams for  the color-octet mechanism 
for $Z^{0} \rightarrow q \bar{q} + J/\psi $ with  $q = u,d,c,s,b$. 
}
\label{figtwo}
\end{figure}

\begin{figure}
\vskip 2cm
\hskip -1cm
\hbox to\textwidth{\hss\epsfig{file=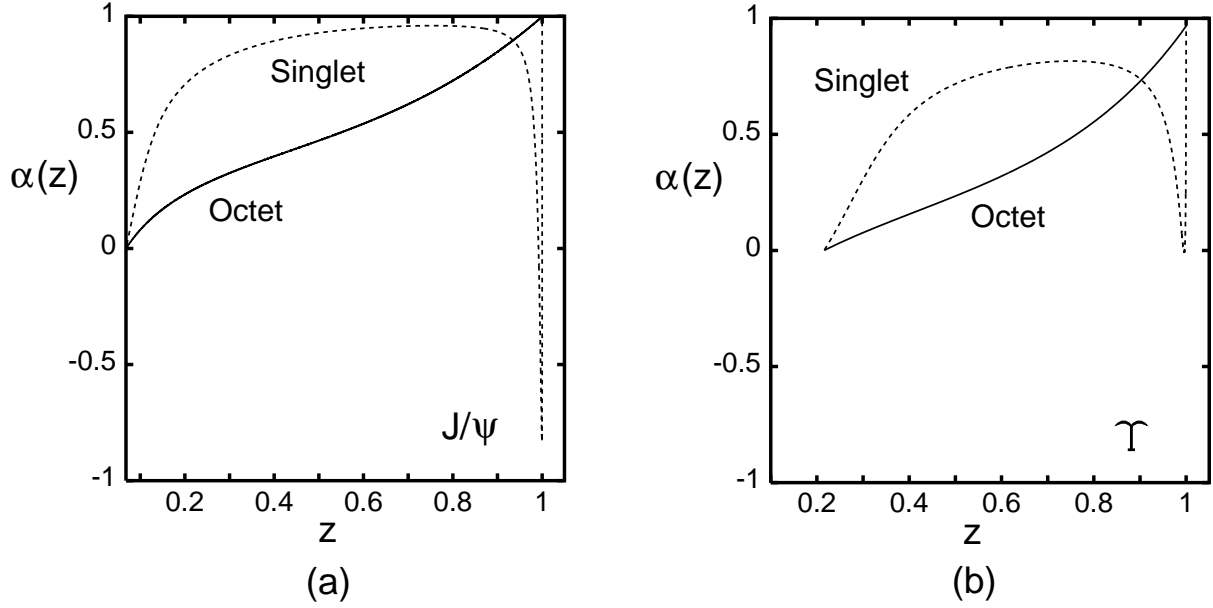,width=16cm}\hss}
\vskip 0.5cm
\caption{ The $\alpha_{1,8} (z)$ of the (a) $J/\psi$ and (b) $\Upsilon$
 as functions of $z \equiv 2 E_H / M_Z~(H=\psi,\Upsilon)$ 
for the color-singlet (in dashed curve) and the color-octet (in solid 
curve) case respectively. }  
\label{figthree}
\end{figure}

\begin{figure}  
\hskip -0.5cm
\hbox to\textwidth{\hss\epsfig{file=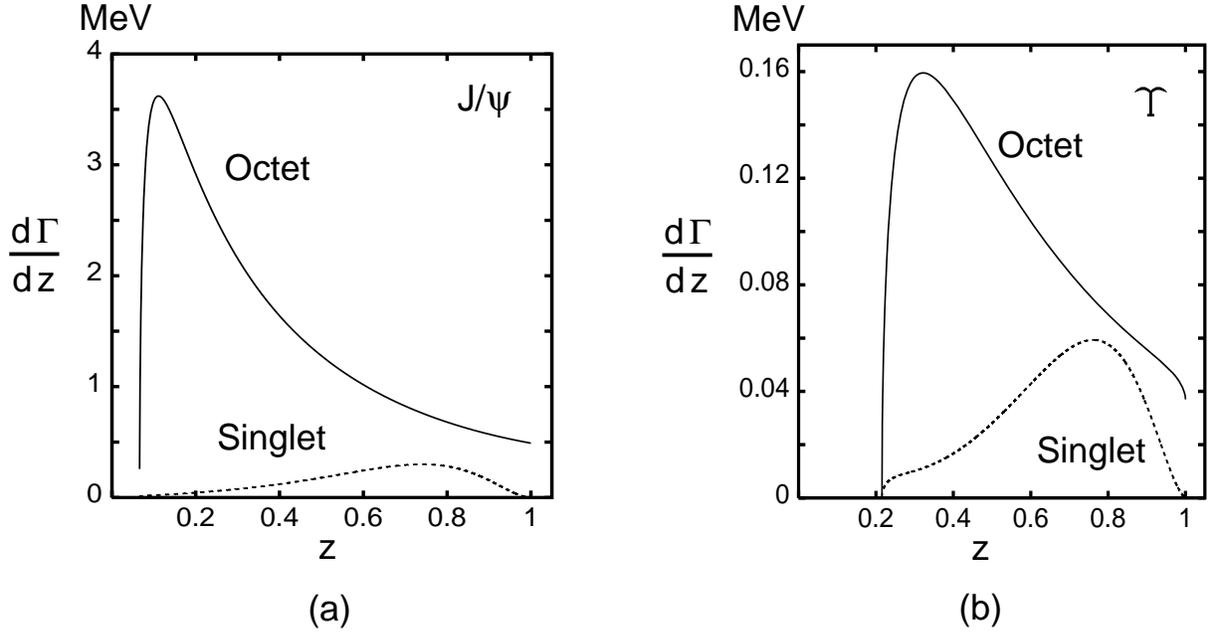,width=16cm}\hss}
\vskip 0.5cm
\caption{The energy  
distributions of the (a) $J/\psi$ and (b) $\Upsilon$.
The color-octet contribution  and the color-singlet contributions are shown
in the solid and the dashed curves, respectively.
}  
\label{figfour}  
\end{figure}

\begin{figure}  
\hbox to\textwidth{\hss\epsfig{file=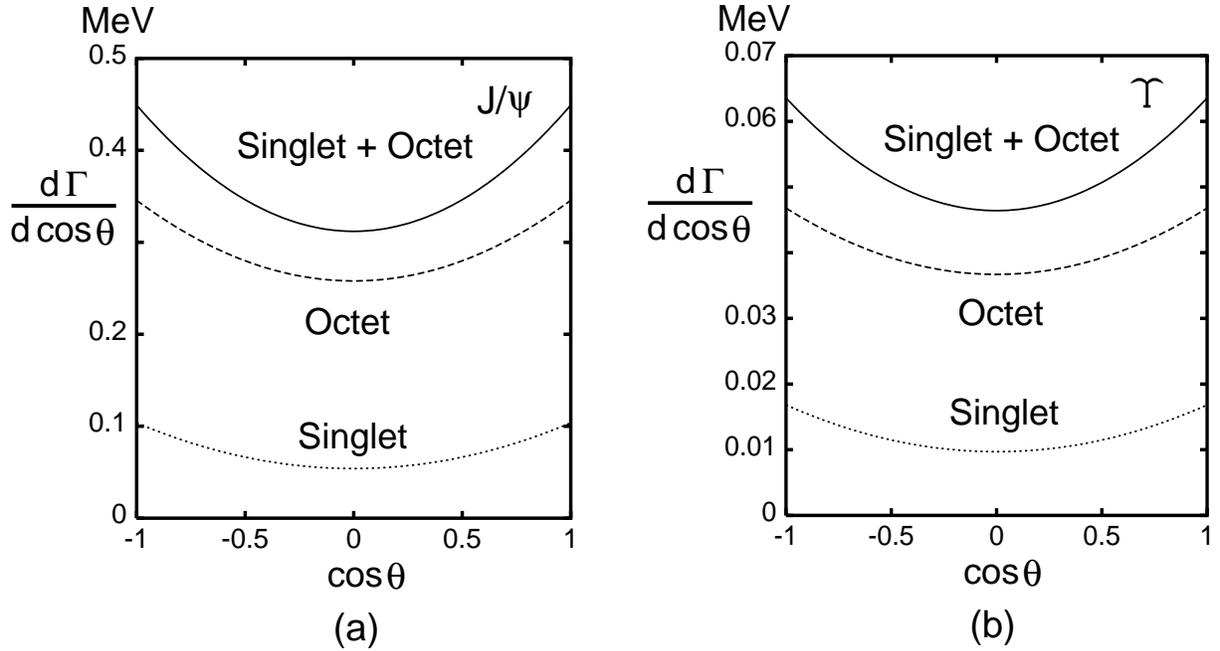,width=16cm}\hss}
\vskip 0.5cm
\caption{
The polar angle
distributions of the (a) $J/\psi$ and (b) $\Upsilon$ (relative to
the $e^{+} e^{-}$ beam direction at the $Z-$peak).
The color-octet contribution, the color-singlet contributions and the sum
of the two are shown
in the dashed, the dotted and the solid curves, respectively.
}  
\label{figfive}  
\end{figure}
\end{document}